\documentclass[prd,amsmath,amssymb,twocolumn]{revtex4-1}

\usepackage{graphicx}
\usepackage{color}
\usepackage{amsmath}
\usepackage{amssymb}

\usepackage{bm}
\usepackage{bbm}

\newcommand{\CH}{\mathbb{C}\otimes\mathbb{H}}
\newcommand{\CO}{\mathbb{C}\otimes\mathbb{O}}
\newcommand{\CHO}{\mathbb{C}\otimes\mathbb{H}\otimes\mathbb{O}}
\newcommand{\RCHO}{\mathbb{R}\otimes\mathbb{C}\otimes\mathbb{H}\otimes\mathbb{O}}
\newcommand{\COOO}{\mathbb{C}\otimes\overleftarrow{\mathbb{O}}}
\newcommand{\CHOOO}{\mathbb{R}\mathbb{C}\overleftarrow{\mathbb{H}{\mathbb{O}}}}
\newcommand{\RCHOOO}{\mathbb{R}\otimes\mathbb{C}\otimes\overleftarrow{\mathbb{H}}\otimes\overleftarrow{\mathbb{O}}}

\newcommand{\CLsix}{\mathbb{C}l(6)}

\newcommand{\C}{\mathbb{C}}
\newcommand{\R}{\mathbb{R}}

\newcommand{\spacer}{0.2cm}
\usepackage{xcolor}

\begin{document}

\title{Three generations, \vspace{.7mm} \\ two unbroken gauge symmetries, \vspace{.7mm} \\  and one eight-dimensional algebra}

\author{N. Furey}
\affiliation{$ $ \\ Department of Applied Mathematics and Theoretical Physics, \\University of Cambridge, \\Wilberforce Road, Cambridge, UK, CB3 0WA\\ and \\ AIMS South Africa, \\6 Melrose Road, Muizenberg, Cape Town, 7945\vspace{2mm}\\  nf252@cam.ac.uk}


\begin{abstract} A considerable amount of the standard model's three-generation structure can be realised from just the $8\hspace{.3mm}\C$-dimensional algebra of the complex octonions.  Indeed, it is a little-known fact that the complex octonions can generate on their own  a $64\hspace{.3mm}\C$-dimensional space.  Here we identify an $su(3)\oplus u(1)$ action which splits this $64\hspace{.3mm}\C$-dimensional space into complexified  generators of $SU(3)$, together with 48 states.   These 48 states exhibit the behaviour of exactly three generations of quarks and leptons under the standard model's two unbroken gauge symmetries.  This article builds on a previous one,~\citep{Gen}, by incorporating electric charge.

Finally, we close this discussion by outlining a proposal for how the standard model's full set of states might be identified within the left action maps of $\RCHO$ (the Clifford algebra $\C l(8)$).  Our aim is to include not only the standard model's three generations of quarks and leptons, but also its gauge bosons.
\end{abstract}

\maketitle

\section{Why three generations?}   

Upon the 2012 discovery of a 125 GeV Higgs, the most straightforward four-generation chiral extension to the standard model was  ruled out, \citep{seal}-\citep{lenz}.  Of course, the possibility of eventually finding a fourth generation is not excluded for every imaginable scenario, e.g. \citep{Htrip}, \citep{multihiggs}.  However, given this new data, it seems increasingly likely that nature's game of replicating particle content comes to an end at three generations.

Although three-generation models can be relatively easy to justify experimentally, they are substantially more difficult to motivate theoretically.  That is, few mathematical objects exhibit (efficiently) the group representations necessary to describe three full generations.  

Indeed, it is no secret that the most well-known extensions of the standard model:  $SU(5)$, $Spin(10)$  grand unified theories, and the Pati-Salam model are all naturally one-generation models.  For the standard model and its most well-studied extensions, the existence of three families need be imposed by hand.

With this being said, a variety of proposals have materialized over the years, e.g. \citep{kong}-\citep{DVtod}, in order to explain the curious pattern.  This includes in particular a recent three-generation proposal put forward by Dubois-Violette and Todorov, based on the 27-dimensional (octonionic) exceptional Jordan algebra.

The intention of our article here is not to present a completed three-generation quantum field theory.   Instead we will demonstrate a significant portion of an algebraic framework on which such a theory might be built.

Rather unconventionally, this framework does not begin with a larger mathematical object, which is then subsequently broken down into the known quark and lepton representations of the standard model.  On the contrary, we will make use of just an $8\hspace{.3mm}\C$-dimensional algebra - an algebra whose degrees of freedom are far outweighed by the number of states which we aim to describe.

We will begin by introducing the algebra of the complex octonions, $\CO$.  This $8\hspace{.3mm}\C$-dimensional algebra will then be seen to generate the complex  Clifford algebra $\CLsix$, via its \it left-action maps. \rm  Within this $64\hspace{.3mm}\C$-dimensional Clifford algebra, we next identify a pair of  complexified $su(3)_c$ Lie algebras.  These  $SU(3)_c$ generators will then be applied to the rest of the Clifford algebra, which consequently breaks down into  exactly  the $SU(3)_c$ representations one would expect for three full generations of quarks and leptons,~\citep{Gen}.  

Subsequently, we demonstrate how the action of these generators may be generalized so as to include a new $u(1)$.  This $U(1)$ action then distributes 48 eigenvalues which are found to coincide with electric charge.  

Hence, it is shown that a single eight-dimensional algebra can  encode the behaviour of three full generations under nature's two unbroken gauge symmetries.

This article builds on~\citep{Gen} by (1) demonstrating that the generators of $G_2$ may be described in terms of \it associators, \rm (2) by redefining the operation of Lie algebras on $\CLsix$ in terms of a single action, (3) by further specifying the projection properties of quark and lepton states, and finally (4) by incorporating electric charge.

Over the years, there has been quite a number of authors who have used $\CLsix$ to describe one generation of standard model fermions.  These include, but are likely not limited to  \citep{Grass}-\citep{gresnigt}.

A natural question which arises in this article is whether or not the demonstrated $SU(3)_c\times U(1)_{em}$ adjoint representations might somehow be identified as gauge bosons.  In this case, one might then hope to eventually describe the standard model's \it full set of states \rm inside of one single mathematical object.  In the addendum of this paper, we outline a proposal for how such a task might be carried out.  Readers wishing to skip straight to this addendum should feel free to do so.

\section{Introduction to $\CO$}  

\begin{figure}[h!]
\begin{center}
\includegraphics[width=6.6cm]{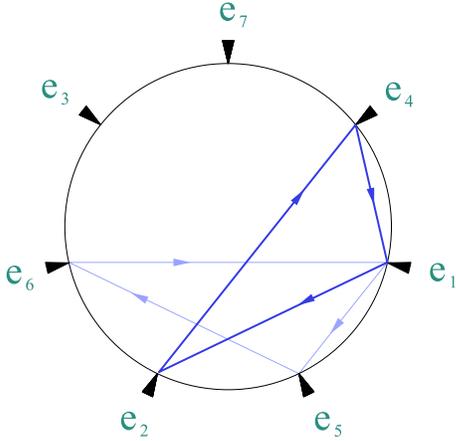}
\caption{\label{circle}
Octonionic multiplication rules,~\citep{GGquarks}.}
\end{center}
\end{figure}

The complex octonions form an $8$-dimensional algebra over $\C$, spanned by basis vectors $e_j$ for $j=0, \dots 7.$  The basis vector $e_0$ plays the role of the multiplicative identity, whereas the $e_k$ for $k=1, \dots 7$ are imaginary units with $e_k^2=-1.$   The remaining octonionic multiplication rules may be described succinctly by specifying $e_1e_2 = e_4$, and then invoking the rules, \citep{baez},
\begin{equation}\begin{array}{l} 
e_ie_j = -e_je_i \hspace{1cm} i\neq j, \vspace{2mm}\\
e_ie_j = e_k \hspace{2mm} \Rightarrow  \hspace{2mm}e_{i+1}e_{j+1} = e_{k+1},\vspace{2mm}\\
e_ie_j = e_k \hspace{2mm}\Rightarrow \hspace{2mm} e_{2i}e_{2j} = e_{2k}. 
\end{array}\end{equation}
\noindent Please see Figure~\ref{circle}.

The octonions are perhaps best known for their property of \it non-associativity, \rm meaning that there exists an  $a,$ $b,$ and $c$ in the algebra such that $a(bc)\neq (ab)c.$  Hence, brackets should typically be specified whenever multiplication involves three or more elements. With this being said, readers should realise that non-associativity is by no means a foreign concept in physics.  It is an under-appreciated fact that both Lie algebras and Jordan algebras likewise constitute non-associative algebras.  (As an example, consider the elements $a=i\sigma_x,$ $b=i\sigma_x,$ $c=i\sigma_y$ in the $su(2)$ Lie algebra, where multiplication is given by the commutator.)


 Symmetries of the algebra:  the derivations  of $\CO$ are given by the complexified $14$-dimensional exceptional Lie algebra $g_2$.   These $g_2$ elements can be seen to act on a generic element $f\in\CO$ as
\begin{equation}\begin{array}{l}  \label{g2}
\Lambda_1 f = \frac{i}{2} \{ e_1,e_5,f \} - \frac{i}{2} \{ e_3,e_4,f \}      \vspace{3mm}   \\

\Lambda_2 f = \frac{i}{2} \{ e_4,e_1,f \} - \frac{i}{2} \{ e_3,e_5,f \} \vspace{3mm}   \\

\Lambda_3 f = -\frac{i}{2} \{ e_1,e_3,f \} + \frac{i}{2} \{ e_4,e_5,f \} \vspace{3mm}   \\

\Lambda_4 f = \frac{i}{2} \{ e_2,e_5,f \} + \frac{i}{2} \{ e_4,e_6,f \} \vspace{3mm}   \\


\Lambda_5 f = -\frac{i}{2} \{ e_2,e_4,f \} + \frac{i}{2} \{ e_5,e_6,f \} \vspace{3mm}   \\

 \Lambda_6 f = \frac{i}{2} \{ e_1,e_6,f \} + \frac{i}{2} \{ e_2,e_3,f \} \vspace{3mm}   \\

  \Lambda_7 f = \frac{i}{2} \{ e_1,e_2,f \} + \frac{i}{2} \{ e_3,e_6,f \} \vspace{3mm}   \\

   \Lambda_8 f = \frac{i}{2\sqrt{3}} \{ e_1,e_3,f \} + \frac{i}{2\sqrt{3}} \{ e_4,e_5,f \} -\frac{i}{\sqrt{3}} \{ e_2,e_6,f \} \vspace{3mm}   \\

   g_9 f = -\frac{i}{2\sqrt{3}} \{ e_1,e_5,f \} - \frac{i}{2\sqrt{3}} \{ e_3,e_4,f \} -\frac{i}{\sqrt{3}} \{ e_2,e_7,f \} \vspace{3mm}   \\

   g_{10} f = -\frac{i}{2\sqrt{3}} \{ e_4,e_1,f \} - \frac{i}{2\sqrt{3}} \{ e_3,e_5,f \} +\frac{i}{\sqrt{3}} \{ e_6,e_7,f \} \vspace{3mm}   \\

  g_{11} f = -\frac{i}{2\sqrt{3}} \{ e_4,e_6,f \} - \frac{i}{2\sqrt{3}} \{ e_5,e_2,f \} +\frac{i}{\sqrt{3}} \{ e_7,e_1,f \} \vspace{3mm}  \\

   g_{12} f = -\frac{i}{2\sqrt{3}} \{ e_2,e_4,f \} - \frac{i}{2\sqrt{3}} \{ e_5,e_6,f \} +\frac{i}{\sqrt{3}} \{ e_3,e_7,f \}  \vspace{3mm}   \\



   g_{13} f = -\frac{i}{2\sqrt{3}} \{ e_6,e_1,f \} - \frac{i}{2\sqrt{3}} \{ e_2,e_3,f \} +\frac{i}{\sqrt{3}} \{ e_7,e_4,f \} \vspace{3mm}   \\

   g_{14} f = -\frac{i}{2\sqrt{3}} \{ e_1,e_2,f \} - \frac{i}{2\sqrt{3}} \{ e_6,e_3,f \} +\frac{i}{\sqrt{3}} \{ e_5,e_7,f \}

\end{array}\end{equation}

\noindent  over $\C$.  Here, we have made use of the \it associator, \rm  defined as $\{a,b,c\} \equiv a(bc)-(ab)c$.   Readers should note that, when taken over the field of the real numbers, the first eight $\Lambda_j$ generate $SU(3)$.  In this case, we have chosen this $SU(3)$ so that it holds the octonionic imaginary unit $e_7$ constant.

\section{From 8 dimensions to 64}

  Given the definition of the associator, it is straightforward to see that the 14 generators of equations~(\ref{g2}) are constructed from chains of octonions acting from the left on $f$.  In fact, the most general \it left-action map, \rm $M,$ may be described as
\begin{equation}\label{6chain}
\begin{array}{ll}
Mf\equiv c_0f  &+ \sum_{i=1}^6 c_i \hspace{.3mm} e_if + \sum_{j=2}^6\sum_{i=1}^{j-1}c_{ij}\hspace{.3mm} e_i(e_jf)  \vspace{3mm}

\\ &+ \sum_{k=3}^6\sum_{j=2}^{k-1}\sum_{i=1}^{j-1}c_{ijk}\hspace{.3mm} e_i(e_j(e_kf)) + \dots  \vspace{3mm}

\\ & + c_{123456}\hspace{.3mm}e_1(e_2(e_3(e_4(e_5(e_6f))))),
\end{array}
\end{equation}
\noindent where the coefficients $c_0,$ $c_i,$ $\dots \in \C$.  Readers may have noticed that the imaginary unit $e_7$  is not explicitly expressed in these maps.  This is due to the fact that 
\begin{equation}\label{e7}
e_7f = e_1(e_2(e_3(e_4(e_5(e_6f))))) \hspace{.5cm} \forall f \in\CO, 
\end{equation}
\noindent thereby making $e_7$ redundant as a left-action map.  Of course, $e_7$ itself holds no preferred status within the octonions, and the space of left-action maps may equivalently be  generated by any six of the seven imaginary units.  

The octonionic chains~(\ref{6chain}) describe all possible complex-linear maps from $\CO$ to itself.  That is, they faithfully represent the full $64\hspace{.3mm}\C$-dimensional space of complex endomorphisms of $\CO$.  Indeed, even right multiplication may be re-expressed in the form of equation~(\ref{6chain}).  For example, 
\begin{equation} \label{e7R} fe_7 = \frac{1}{2}\left( e_1(e_3f) + e_2(e_6f) + e_4(e_5f) - e_7f \right)
\end{equation}
\noindent $\forall f\in \CO$.

So in summary, we have shown how it is possible to build up a $64\hspace{.3mm}\C$-dimensional space, using only the $8\hspace{.3mm}\C$-dimensional $\CO$ operating on itself from the left.  For examples of earlier work which make reference to this $64$-dimensional algebra, see \citep{Grass}, \citep{deleo}, \citep{dixon_fam}.

\section{The Clifford algebra $\CLsix$}

  It is reasonably straightforward to show that
\begin{equation} 
e_i(e_jf ) =      
\begin{cases}\hspace{2mm} -e_j(e_if) \hspace{6mm} \textup{when } i\neq j \vspace{3mm} \\

		\hspace{2mm}	-f \hspace{14.5mm} \textup{when } i= j
\end{cases}			
\end{equation}
\noindent $\forall f\in \CO,$ and for $i,j = 1,\dots 6.$  (This should be recognizable to the reader as Clifford algebraic structure.)  In fact, the linear maps~(\ref{6chain}) give a faithful representation of the complex Clifford algebra $\CLsix.$

Readers concerned about the potential conflict between the inherent associativity of a Clifford algebra, and the  non-associativity of the octonions should note that \it multiplication \rm of left-action maps is given by the  \it composition of maps. \rm  Of course, the composition of maps is  associative, by definition; $F\circ( G \circ H) = (F\circ G) \circ H$.

Consider now for a moment complex Clifford algebras of the form $\C l(2n)$ for $n\in \mathbb{Z}>0$.  It is known that the generating space of such algebras may be partitioned into an $n$-dimensional subspace spanned by raising operators $\{\alpha_i^{\dagger}\}$ and an $n$-dimensional subspace spanned by lowering operators $\{\alpha_i\}$.  These $n$-dimensional subspaces are known as \it maximal totally isotropic subspaces, \rm whose basis vectors obey
\begin{equation}\label{anticomm}\{\alpha_i, \alpha_j\}=0=\{\alpha_i^{\dagger}, \alpha_j^{\dagger}\}, \hspace{6mm} \{\alpha_i, \alpha_j^{\dagger}\}=\delta_{ij}
\end{equation}
\noindent for $i,j=1,\dots n$, under the anticommutator: $\{a,b\}\equiv ab+ba$.  For further details, please see~\citep{ablam}.

In the case of our octonionic representation of $\CLsix$, (\ref{6chain}), the generating space is spanned by the linear maps $e_i$ for $i=1,\dots 6$.  These may be reorganized into a set of lowering operators
\begin{equation} \alpha_1 \equiv \frac{-e_5 +ie_4}{2}, \hspace{3mm} \alpha_2 \equiv \frac{-e_3 +ie_1}{2}, \hspace{3mm} \alpha_3 \equiv \frac{-e_6 +ie_2}{2},
\end{equation}
\noindent and a set of raising operators, 
\begin{equation} \alpha_1^{\dagger} \equiv \frac{e_5 +ie_4}{2}, \hspace{3mm} \alpha_2^{\dagger} \equiv \frac{e_3 +ie_1}{2}, \hspace{3mm} \alpha_3^{\dagger} \equiv \frac{e_6 +ie_2}{2},
\end{equation}
\noindent where $\dagger$ maps the complex $i\mapsto -i$ and the octonionic $e_j\mapsto -e_j$, while reversing the order of multiplication, $(ab)^{\dagger} = b^{\dagger} a^{\dagger}$.  Readers may confirm that these ladder operators obey equations~(\ref{anticomm}), as maps acting on any $f\in \CO.$

\begin{figure}[h!]
\begin{center}
\vspace{5mm}
\includegraphics[width=7cm]{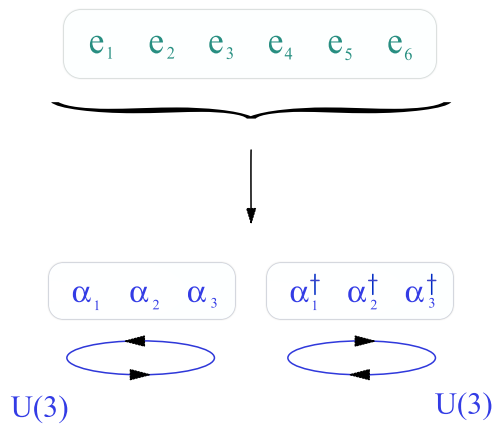}
\caption{\label{cob}   Octonionic imaginary units $e_1, e_2,\dots e_6$ generate the Clifford algebra $\CLsix$.  These six generators may be rewritten in terms of a basis of ladder operators, $\alpha_1, \alpha_2, \alpha_3, \alpha_1^{\dagger}, \alpha_2^{\dagger}, \alpha_3^{\dagger},$ which have $U(3) = SU(3)\times U(1)/ \mathbb{Z}_3$ symmetry.}
\end{center}
\end{figure}

The structure of these ladder operators is preserved by the unitary group $U(3)=SU(3)\times U(1) / \mathbb{Z}_3$, as depicted in Figure~(\ref{cob}).  
\noindent This $U(3)$ symmetry may be realised as $G\alpha_iG^{-1}$ and $G\alpha_i^{\dagger}G^{-1}$, for $G\equiv \exp{\left(\hspace{.5mm} ir_j\Lambda_j + ir_0Q \hspace{.5mm}\right)}\in\CLsix$.  Here, $r_k\in \R$ for $k=0, \dots 8$, and the $SU(3)$ generators, $\Lambda_j$, are defined as in equations~(\ref{g2}).  The $U(1)$ generator is defined as
\begin{equation}\label{Q}Qf\equiv \frac{N}{3}f = \frac{1}{3} \sum_{i=1}^3 \alpha_i^{\dagger}\left(\alpha_if\right),
\end{equation}
\noindent  acting on any $f\in\CO$.    This $Q$ is proportional to the number operator, $N$, for the system, and can be seen to commute with the $\Lambda_j$.   In~\citep{qq}, $Q$ was identified as the generator of \it electric charge, \rm in the context of a one-generation model.

Before moving on, we will first simplify our notation.  From here forward,  it will now be implicitly assumed that equations in $\CLsix$ hold $\forall f\in\CO.$  That is, we will no longer write $f$ explicitly.  Furthermore, we will cease to write the nested brackets of octonionic left-action maps, (\ref{6chain}).  That is, right-to-left bracketing will also be implicitly assumed.  Hence, equations such as $Qf\equiv  \frac{1}{3} \sum_{i=1}^3 \alpha_i^{\dagger}\left(\alpha_if\right)$ will now simply read $Q\equiv  \frac{1}{3} \sum_{i=1}^3 \alpha_i^{\dagger}\alpha_i.$

\section{Three generations under $SU(3)_c$}

 For our first result, we will now show how the $SU(3)_c$ irreducible representations corresponding to three generations of quarks and leptons may be found, using only the action of  $\CO$ on itself.  

We begin by splitting $\CLsix$ into two $32\hspace{.3mm}\C$-dimensional pieces:  $\CLsix s$ and $\CLsix s^*$, where $s$ is given by the linear map $s\equiv \frac{1}{2}\hspace{.5mm}(1+ie_7)\in\CLsix$.  Readers may confirm that both $s$ and $s^*$ are idempotents, and that   $ss^*=s^*s=0.$  

Within the subalgebra $\CLsix s$, we find a faithful representation of the Lie algebra $su(3)$, generated by eight objects of the form $\Lambda_js$.  Seeing as how $\left[\hspace{.5mm} \Lambda_j, \hspace{.5mm} s\hspace{.5mm} \right]=0,$ it may be confirmed that 
\begin{equation}   \left[ \hspace{.5mm} \frac{\Lambda_i}{2}s, \hspace{.5mm} \frac{\Lambda_j}{2}s \hspace{.5mm}\right] = i c_{ijk} \frac{\Lambda_k}{2}s
\end{equation}
\noindent holds, where $c_{ijk}$ are the usual $SU(3)$ structure constants.

Now, given this representation of the $SU(3)$ Lie algebra, we may subsequenty apply the $\Lambda_js$ generators to the remainder of $\CLsix s.$  Under the action $[\hspace{.5mm}i\Lambda_js, \hspace{.5mm}\CLsix s\hspace{.5mm}],$ the $32\hspace{.3mm}\C$-dimensional  $\CLsix s$ is found to break  down as
\begin{equation}\label{32break}
 \CLsix s \hspace{2.8mm}\mapsto \hspace{2.8mm}\mathbf{\underline{8}}\hspace{.8mm} \oplus \hspace{.8mm}\mathbf{\underline{3}} \hspace{.8mm}\oplus \hspace{.8mm}\left( 5 \times \mathbf{\underline{3}^*}\right) \hspace{.7mm}\oplus \hspace{.7mm}\left( 6 \times \mathbf{\underline{1}}\right)
\end{equation}
\noindent over $\C.$  For a sample calculation, please see~\citep{Gen}.  Invoking the complex conjugate, $i\mapsto -i$, sends particles to antiparticles, and vice versa.  In other words, the commutator $[\hspace{.5mm}-i\Lambda_j^*s^*, \hspace{.5mm}\CLsix s^*\hspace{.5mm}]$ induces 
\begin{equation}\label{32break_star}
 \CLsix s^* \hspace{2.8mm}\mapsto \hspace{2.8mm}\mathbf{\underline{8}}\hspace{.8mm} \oplus \hspace{.8mm}\mathbf{\underline{3}}^* \hspace{.8mm}\oplus \hspace{.8mm}\left( 5 \times \mathbf{\underline{3}}\right) \hspace{.7mm}\oplus \hspace{.7mm}\left( 6 \times \mathbf{\underline{1}}\right).
\end{equation}

Finally, these actions may be trivially combined into one single action on the full $64\hspace{.3mm}\C$-dimensional $\CLsix.$  Under $[\hspace{.5mm}i\Lambda_js, \hspace{.5mm}\CLsix s\hspace{.5mm}] + [\hspace{.5mm}-i\Lambda_j^*s^*, \hspace{.5mm}\CLsix s^*\hspace{.5mm}], $ the algebra $\CLsix$ breaks down into a pair of complexified $SU(3)$ Lie algebras, together with the $SU(3)$ representations
\begin{equation}\label{64break}
\left( 6 \times \mathbf{\underline{3}}\right) \hspace{.7mm} \oplus \hspace{.7mm}\left( 6 \times \mathbf{\underline{3}}^*\right) \hspace{.7mm} \oplus \hspace{.7mm} \left( 6 \times \mathbf{\underline{1}} \right) \hspace{.7mm} \oplus \hspace{.7mm} \left( 6 \times \mathbf{\underline{1}}\right).
\end{equation}
\noindent Readers should recognize these as the $SU(3)_c$ representations necessary to describe three full generations of quarks and leptons.

\section{Three generations under $SU(3)_c\times U(1)_{em}/\mathbb{Z}_3$}

  Now, given the ladder operator symmetry $U(3) = SU(3)_c\times U(1)_{em}/\mathbb{Z}_3$ described earlier, it is natural to wonder if these $SU(3)_c$ results may be extended so as to include $U(1)_{em}$.  The most obvious electromagnetic extension to the action $[\hspace{.5mm}i\Lambda_js, \hspace{.5mm}\CLsix s\hspace{.5mm}]+ [\hspace{.5mm}-i\Lambda_j^*s^*, \hspace{.5mm}\CLsix s^*\hspace{.5mm}] $ is clearly $[\hspace{.5mm}iQs, \hspace{.5mm}\CLsix s\hspace{.5mm}] + [\hspace{.5mm}-iQ^*s^*, \hspace{.5mm}\CLsix s^*\hspace{.5mm}].$  However, we find that this action fails to assign the correct electric charges.  

Hence, we are then left to ask:  \it Could there be a way to generalize this action so that $Q$ produces  the electric charges of standard model fermions? \rm

In what is to follow, we will need to introduce an idempotent, $S$, which is the right-multiplication analogue of $s$.  
\begin{equation}\begin{array}{ll}\label{S} Sf  & \equiv f\frac{1}{2}(1+ie_7) 
\vspace{2mm}
\\ &= \frac{1}{2}f -\frac{i}{4}e_7f+\frac{i}{4}e_1(e_3f)+\frac{i}{4}e_2(e_6f)+\frac{i}{4}e_4(e_5f)  \vspace{1mm}
\end{array}\end{equation}
\noindent acting on $f\in\CO,$ or more simply, 
\begin{equation}\label{Scompact}
S\equiv \frac{1}{2}  + \frac{1}{4}\left(  -ie_7+ie_{13}+ie_{26}+ie_{45}\right),
\end{equation} 
\noindent where $e_{ab}f$ is shorthand for $e_a(e_bf).$  Readers will find that equation~(\ref{S}) is easily confirmed given equation~(\ref{e7R}).  As before, $SS^* = S^*S=0,$ and furthermore, $\left[\hspace{.5mm}s,S\hspace{.5mm}\right] = \left[\hspace{.5mm}s, S^*\hspace{.3mm}\right]=0.$

Given that $S+S^*=1$, it is straightforward to see that the action $[\hspace{.5mm}i\Lambda_js, \hspace{.5mm}\CLsix s\hspace{.5mm}]+ [\hspace{.5mm}-i\Lambda_j^*s^*, \hspace{.5mm}\CLsix s^*\hspace{.5mm}] $ is equal to 
\begin{equation}\begin{array}{l} \label{newaction} 
\left[\hspace{1mm}i\Lambda_j s,\hspace{1mm} S\hspace{.5mm}\CLsix\hspace{.2mm} s \hspace{1mm}\right]
+\left[-i\Lambda_j^* s^*, \hspace{1mm}S^*\hspace{.2mm}\CLsix\hspace{.2mm} s^* \hspace{1mm}\right]   \vspace{2mm}\\
+\left[\hspace{1mm}i\Lambda_j s,\hspace{1mm} S^*\hspace{.2mm}\CLsix\hspace{.2mm} s \hspace{1mm}\right] 
+\left[-i\Lambda_j^* s^*, \hspace{1mm}S\hspace{.5mm}\CLsix \hspace{.2mm}s^* \hspace{1mm}\right].
\end{array}\end{equation}
\noindent Furthermore, from equations~(\ref{g2}), it is clear that $\Lambda_j^* = -\Lambda_j$, and hence the action~(\ref{newaction}) is identical to
\begin{equation}\begin{array}{l} \label{newaction_swap} 
\left[-i\Lambda_j^* s,\hspace{1mm} S\hspace{.5mm}\CLsix\hspace{.2mm} s \hspace{1mm}\right]
+\left[\hspace{1mm}i\Lambda_j s^*, \hspace{1mm}S^*\hspace{.2mm}\CLsix\hspace{.2mm} s^* \hspace{1mm}\right]   \vspace{2mm}\\
+\left[\hspace{1mm}i\Lambda_j s,\hspace{1mm} S^*\hspace{.2mm}\CLsix\hspace{.2mm} s \hspace{1mm}\right] 
+\left[-i\Lambda_j^* s^*, \hspace{1mm}S\hspace{.5mm}\CLsix \hspace{.2mm}s^* \hspace{1mm}\right].
\end{array}\end{equation}
\noindent However, this new action is not the same as the old action when we extend these generators so as to include  $Q$.  Unlike with the $\Lambda_j$ operators, $Q^*\neq -Q.$

\noindent Upon finally including $Q,$ 
\begin{equation}\begin{array}{l} \label{newaction_full} 
\left[-i(r_j\Lambda_j^*+r_0Q^* )s,\hspace{1mm} S\hspace{.5mm}\CLsix\hspace{.5mm} s \hspace{1mm}\right]  \vspace{2mm}\\
\hspace{5mm}+\left[\hspace{1mm}i(r_j\Lambda_j+r_0Q) s^*, \hspace{1mm}S^*\hspace{.5mm}\CLsix\hspace{.5mm} s^* \hspace{1mm}\right]   \vspace{2mm}\\
\hspace{10mm}+\left[\hspace{1mm}i(r_j\Lambda_j+r_0Q) s,\hspace{1mm} S^*\hspace{.5mm}\CLsix \hspace{.5mm}s \hspace{1mm}\right]  \vspace{2mm}\\
\hspace{15mm}+\left[-i(r_j\Lambda_j^*+r_0Q^*) s^*, \hspace{1mm}S\hspace{.5mm}\CLsix \hspace{.5mm}s^* \hspace{1mm}\right],
\end{array}\end{equation}
\noindent we find that $\CLsix$ now breaks down as
\begin{equation}\label{fullbreakdown}\begin{array}{l}
 \CLsix  \hspace{0.8mm}\mapsto \vspace{2.5mm}\\
 
 \mathbf{\underline{8}_{ \hspace{.6mm}0}}\hspace{.7mm} 
 \oplus \hspace{.7mm} (3\times\mathbf{\underline{3}_{ \hspace{.6mm}\frac{2}{3}}}  )
 \hspace{.7mm}\oplus \hspace{.7mm} ( 3\times\mathbf{\underline{3}_{ -\frac{1}{3}}} ) \hspace{.7mm}\oplus \hspace{.7mm} ( 3 \times \mathbf{\underline{1}_{\hspace{.6mm}0}} )\hspace{.7mm}\oplus \hspace{.7mm} ( 3 \times \mathbf{\underline{1}_{-1}} )\hspace{.7mm}\oplus \vspace{2mm}\\
 
  \mathbf{\underline{8}_{ \hspace{.6mm}0}}\hspace{.7mm} 
 \oplus \hspace{.7mm} (3\times\mathbf{\underline{3}^*_{\hspace{.1mm} -\frac{2}{3}}}  )
 \hspace{.7mm}\oplus \hspace{.7mm} ( 3\times\mathbf{\underline{3}^*_{ \hspace{.7mm}\frac{1}{3}}} ) \hspace{.7mm}\oplus \hspace{.7mm} ( 3 \times \mathbf{\underline{1}_{\hspace{.6mm}0}} )\hspace{.7mm}\oplus \hspace{.7mm} ( 3 \times \mathbf{\underline{1}_{\hspace{.6mm}1}} ).\vspace{3mm}
\end{array}\end{equation}

\vspace{2mm}

To be more explicit,  the $SU(3)_c\times U(1)_{em}/\mathbb{Z}_3$ representations corresponding to three generations of particles may be described by complex linear combinations of the states
$$\begin{array}{l}\label{u}
 \color{black} \mathbf{\underline{3}_{  \hspace{.6mm} \frac{2}{3}}}\color{black}   \hspace{1mm} \rightarrow \hspace{1mm}
  \begin{cases}
  
\hspace{3mm} \color{black} u_1^R \color{black}  \equiv   sS^*\left(-ie_{12}-e_{16}+e_{23}+ie_{36}\right)sS    \vspace{\spacer} \\
\hspace{3mm} \color{black} u_1^G \color{black}\equiv   sS^*\left(-ie_{24}-e_{25}+e_{46}-ie_{56}\right)sS \vspace{\spacer} \\
\hspace{3mm} \color{black} u_1^B\color{black}\equiv    sS^*\left(ie_{14}+e_{15}+e_{34}-ie_{35}\right)sS 
\end{cases}\vspace{3mm}\\

 \color{black} \mathbf{\underline{3}_{  \hspace{.6mm} \frac{2}{3}}}\color{black}  \hspace{1mm} \rightarrow \hspace{1mm}
  \begin{cases}
\hspace{3mm} \color{black}{u}_2^R\color{black}\equiv    s^*S^* \left(-ie_{12}-e_{16}+e_{23}+ie_{36}\right)s^*S    \vspace{\spacer} \\
\hspace{3mm} \color{black}{u}_2^G\color{black}\equiv   s^*S^* \left(-ie_{24}-e_{25}+e_{46}-ie_{56}\right)s^*S  \vspace{\spacer} \\
\hspace{3mm} \color{black}{u}_2^B\color{black}\equiv   s^*S^* \left(ie_{14}+e_{15}+e_{34}-ie_{35}\right)s^*S
\end{cases}\vspace{3mm}\\

 \color{black} \mathbf{\underline{3}_{  \hspace{.6mm} \frac{2}{3}}} \color{black} \hspace{1mm} \rightarrow \hspace{1mm}
  \begin{cases}
\hspace{3mm} \color{black}{u}_3^R\color{black}\equiv  sS \left(-ie_{4}+e_{5}+e_{134}+ie_{135}\right)s^*S  \vspace{\spacer} \\
\hspace{3mm} \color{black}{u}_3^G\color{black}\equiv   sS \left(-ie_{1}+e_{3}+e_{126}+e_{145}\right)s^* S \vspace{\spacer} \\
\hspace{3mm} \color{black}{u}_3^B\color{black}\equiv  sS \left(-ie_{2}+e_{6}-e_{123}+ie_{136}\right)s^*S
\end{cases} 
\end{array}
$$

$$\begin{array}{l}\label{d}
  \color{black}\mathbf{\underline{3}_{  \hspace{.2mm} -\frac{1}{3}}} \color{black} \hspace{1mm} \rightarrow \hspace{1mm}
  \begin{cases}
\hspace{3mm} \color{black}{d}_1^R\color{black}\equiv       sS^*\left(-ie_{1}-e_{3}+e_{126}-e_{145}\right)s^*S  \vspace{\spacer} \\
\hspace{3mm} \color{black}{d}_1^G\color{black}\equiv        sS^*\left(ie_{4}+e_{5}+e_{134}-ie_{135}\right)s^*  S  \vspace{\spacer} \\
\hspace{3mm} \color{black}{d}_1^B\color{black}\equiv         sS^*\left(-ie_{124}-e_{125}-e_{146}+ie_{156}\right)s^* S
\end{cases} \vspace{3mm}\\

 \color{black} \mathbf{\underline{3}_{  \hspace{.2mm} -\frac{1}{3}}}\color{black}  \hspace{1mm} \rightarrow \hspace{1mm}
  \begin{cases}

\hspace{3mm} \color{black}{d}_2^R\color{black}\equiv    sS^*  \left(ie_{2}+e_{6}+e_{123}+ie_{136}\right)s^* S  \vspace{\spacer} \\
\hspace{3mm} \color{black}{d}_2^G\color{black}\equiv    s  S^*   \left(-ie_{124}-e_{125}+e_{146}-ie_{156}\right)s^* S   \vspace{\spacer} \\
\hspace{3mm} \color{black}{d}_2^B\color{black}\equiv      s S^*   \left(-ie_{4}-e_{5}+e_{134}-ie_{135}\right)s^*S  
\end{cases} \vspace{3mm}\\

 \color{black} \mathbf{\underline{3}_{  \hspace{.2mm} -\frac{1}{3}}} \color{black} \hspace{1mm} \rightarrow \hspace{1mm}
  \begin{cases}

\hspace{3mm} \color{black}{d}_3^R \color{black}   \equiv          s S^*\left(-ie_{124}+e_{125}+e_{146}+ie_{156}\right)s^* S  \vspace{\spacer} \\
\hspace{3mm} \color{black}{d}_3^G  \color{black}   \equiv       s S^*\left(-ie_{2}-e_{6}+e_{123}+ie_{136}\right)s^*  S  \vspace{\spacer} \\
\hspace{3mm} \color{black}{d}_3^B   \color{black}   \equiv      s  S^*\left(ie_{1}+e_{3}+e_{126}-e_{145}\right)s^*S

\end{cases} 

\end{array}$$

$$\begin{array}{l}\label{n}
 \color{black}  \mathbf{\underline{1}_{  \hspace{.7mm} 0}} \color{black} \hspace{4mm} \rightarrow \hspace{4mm} \color{black}\nu_1\color{black}\equiv       sS\left(1+ie_{13}+ie_{26}+ie_{45}\right)s S       \vspace{3mm} \\

 \color{black}\mathbf{\underline{1}_{  \hspace{.7mm} 0}}  \color{black}\hspace{4mm} \rightarrow \hspace{4mm} \color{black}\nu_2\color{black}\equiv      s S^*  \left(3-ie_{13}-ie_{26}-ie_{45}\right)s S^*
\vspace{3mm} \\

 \color{black}\mathbf{\underline{1}_{  \hspace{.7mm} 0}} \color{black} \hspace{4mm} \rightarrow \hspace{4mm} \color{black}\nu_3\color{black}\equiv    s^*S^* \left(-ie_{124}-e_{125}+e_{146}-ie_{156}\right)sS

\end{array}$$

\begin{equation}\begin{array}{l}\label{particles}
 \color{black}\mathbf{\underline{1}_{  \hspace{.2mm} -1}}\color{black}  \hspace{4mm} \rightarrow \hspace{4mm} \color{black}e^-_1\color{black}\equiv     s S^* \left(ie_{1}-e_{3}+e_{126}+e_{145}\right)s^*  S^*    \vspace{3mm} \\

 \color{black}\mathbf{\underline{1}_{  \hspace{.2mm} -1}} \color{black} \hspace{4mm} \rightarrow \hspace{4mm} \color{black}e^-_2\color{black}\equiv   s S^*\left(-ie_{2}+e_{6}+e_{123}-ie_{136}\right)s^*S^*    \vspace{3mm} \\

 \color{black}\mathbf{\underline{1}_{  \hspace{.2mm} -1}} \color{black} \hspace{4mm} \rightarrow \hspace{4mm} \color{black}e^-_3\color{black}\equiv  s S^* \left(-ie_{4}+e_{5}-e_{134}-ie_{135}\right)s^*S^*. \vspace{2mm}
\end{array}\end{equation}
\noindent From here, finding anti-particle states is remarkably easy.  As with previous work,  \citep{GGquarks}, \citep{Gen}, \citep{qq}, \citep{thesis}, one simply invokes the complex conjugate, $i\mapsto -i.$  

Readers should note that we are not distinguishing between the generations at this point.  Hence, for example, the three $\mathbf{\underline{3}_{  \hspace{.6mm} \frac{2}{3}}}$ representations are labeled arbitrarily as $u_1,$ $u_2,$ $u_3,$ instead of $u,$ $c,$ $t$.

Finally, we mention that the electric charge assignments of equations~(\ref{particles}) can easily be confirmed by the reader.  This is facilitated by the fact that $Q$ may be decomposed as
\begin{equation}\label{Qproj} Q = \frac{1}{3}\hspace{.5mm}s^*S+\frac{2}{3}\hspace{.5mm}sS^*+s^*S^*.
\end{equation}
\noindent For example, $Q$ may be applied to the state $\color{black}u_1^R\color{black}$ by setting $r_j=0$ and $r_0=1$ in the action~(\ref{newaction_full}): 
\begin{equation}\begin{array}{ll}\left[\hspace{.5mm}i\hspace{.5mm}Qs, \hspace{.5mm}S^*\hspace{.2mm}\color{black}u_1^R\color{black}\hspace{.2mm}s\hspace{.5mm}\right] &= i \left[\hspace{.7mm}\frac{2}{3}\hspace{.5mm}sS^*, \hspace{.5mm}sS^*\hspace{.2mm}\color{black}u_1^R\color{black}\hspace{.2mm}sS\hspace{.7mm}\right] \vspace{3mm}\\

&= i\hspace{.9mm}\frac{2}{3}\hspace{.2mm}\color{black}u_1^R\color{black}.
\end{array}\end{equation}
\noindent Hence, the action~(\ref{newaction_full}) assigns to $\color{black}u_1^R\color{black}$ a $Q$ charge of  $\frac{2}{3}$.

\section{Summary}

  This article demonstrates how the $SU(3)_c\times U(1)_{em}/\mathbb{Z}_3$ representations for three full generations of quarks and leptons may be generated, using just an $8\hspace{.3mm}\C$-dimensional algebra.  In order to arrive at these 48 states, we did not simply replicate copies of $\CO$.  Instead, we considered the action of this one  algebra on itself. 

For those more accustomed to grand unified theories, this method should indeed seem unfamiliar.  That is, it runs backwards to the usual direction of the prototypical unified theory.   Standard grand unified theories begin with a sizeable Lie group, and then implement an appropriate mechanism in order to scale the symmetry group down.  In contrast, this article shows how a low-dimensional algebra may act autonomously in order to scale the degrees of freedom up.

Although we have not proposed a grand unified theory here, these results do seem to point towards unification of another form.  It is clear that the $SU(3)_c\times U(1)_{em}/\mathbb{Z}_3$ group elements, and also these 48 states, owe their existence to the same algebra.  Ideally, \it all \rm objects in such a model should likewise arise from the same algebra.

\vspace{.2cm}

\section{Outlook}

  While $\CO$ did supply a reasonable portion of the standard model's group representation structure, we have by no means achieved a full description.  For instance, nowhere in this paper have we discussed spin or chirality.  And so we ask, \it in what ways may these results be extended? \rm

In the third chapter of \citep{thesis}, it was shown that each of the Lorentz representations of the standard model can be identified as invariant subspaces of the algebra of the complex quaternions, $\CH.$  To be more precise, this $4\hspace{.3mm}\C$-dimensional algebra yields: Lorentz scalars, $\phi$, left- and right-handed Weyl spinors, $\Psi_L, \hspace{.5mm}\Psi_R$, Majorana spinors, $\Psi_M$, Dirac spinors, $\Psi_D,$ four-vectors, $p_{\mu},$ and the field strength tensor, $F_{\mu\nu}$, \citep{conway}.  These Lorentz  (or $SL(2,\C)$) representations were identified as invariant subspaces of $\CH$ under various actions of the algebra on itself.  In each case, they were found to arise as a result of the outer automorphism and anti-automorphisms of the algebra, \citep{thesis}.


It is straightforward to see that $\CH$ and $\CO$ may be combined, via a tensor product over $\C$, into the algebra $\left(\CH\right)\otimes_{\C}\left(\CO\right) = \CHO = \RCHO$.  The Dixon algebra $\RCHO$ is the tensor product of the \it only \rm four normed division algebras over the real numbers.  (For an alternate  three-generation model which makes use of  tensor products of column vectors over division algebras, see~\citep{dixon_fam}.)

Once spin and chirality are incorporated into this model, we might then be in a position to address some obvious outstanding questions.  For example, 
\vspace{2mm}

\noindent $\hspace{1mm}\cdot\hspace{1mm}$\it  How is electroweak symmetry to be described in this model?   How might it incorporate $Q$?

\noindent $\hspace{1mm}\cdot\hspace{1mm}$\it How do we interpret the $\hspace{.7mm} \mathbf{\underline{8}_{  \hspace{.7mm} 0}}$ representations in $\CLsix$? 

\noindent $\hspace{1mm}\cdot\hspace{1mm}$\it What brings about the  form of the action~(\ref{newaction_full})?

\noindent $\hspace{1mm}\cdot\hspace{1mm}$\it What is the connection between the recent one-generation results of~\citep{malala} and this model?  Or should~\citep{malala} prompt the reconsideration of a 4-generation model?

\noindent $\hspace{1mm}\cdot\hspace{1mm}$\it If the standard model's group representation structure is indeed a result of the algebras $\R$, $\C$, $\mathbb{H}$, and $\mathbb{O},$ then what is it exactly that is so special about these algebras?  \rm

\section{Addendum:  \\Towards a complete description}

In recent months there has been quite a number of papers written, which link $\R$, $\C$, $\mathbb{H}$, and $\mathbb{O}$ to elementary particle physics.  Notably, Gording and Schmidt-May~\cite{gsm} have carefully demonstrated that the full set of standard model bosons and fermions can be neatly accounted for within $\COOO$, supplemented by the complex quaternions.   Dubois-Violette and Todorov continue the pursuit of linking the exceptional Jordan algebra to the standard model~\cite{dvt2019}; a broad series of exploratory papers has been written by Castro-Perelman, \cite{carlos_RCHOgut}-\cite{carlos_3gen}, including a new four-generation model.  Gresnigt and Gillard have proposed a description of three generations of fermions in terms of $\C l(8)$ with possible links to triality, [41].  Also closely related is a carefully crafted article by Zatloukal, which advocates for a Clifford algebraic description of non-abelian gauge theories, \cite{zat}.  

Given this recent activity, we would like to contribute a brief proposal, which outlines some new ideas for further investigation.  This \it preliminary \rm proposal  is aimed at uncovering a novel  algebraic description of the standard model's \it full set of states.  \rm It extends ideas from the open questions section of \cite{Gen}. 

In short, we offer the idea that the standard model's states might arise as the result of a \it multi-action. \rm  A multi-action is a new kind of generalized multiplication rule, which single-handedly splits a Clifford algebra into Lie algebras, Jordan algebras, and spinor spaces. 

Suppose we were to consider the Clifford algebra $\C l(8)$, generated now by left multiplicative chains of $\RCHO$.  Gording and Schmidt-May recently demonstrated in detail, for the first time, that this $\RCHOOO$ with its $256$ $\C$ dimensions has the capacity to house these particle representations.  This would include not only the standard model's three generations of fermions, but also its bosons.  Unlike with the work presented earlier in this paper, these states could now finally include spin and polarization.

\subsection{Set up}

In seeking out a final unified algebraic description, we might hope for a couple of features to materialize.

\bf(1) \rm  We might expect that the theory's chirality be an artifact of some natural algebraic structure, and will aim to avoid the use of an ad hoc projection operator for the task.  

\bf(2) \rm We might similarly hope that those states describing the three different fermion generations will be found to be algebraically distinct from one another (not just three copies of one another).   This may prove to be a valuable condition for a model which ultimately describes the fermion mass spectrum.

\vspace{2mm}

Our goal now is to describe the standard model's full set of states, in its entirety.  The machinery at our disposal consists of a 256 $\C$-dimensional incarnation of $\C l(8),$ generated by $\CHO$.  The question is then:  \it How should this algebra act on itself so that it splinters into all the familiar elementary particle representations? \rm  Should the algebra act via left multiplication?  Lie bracket?  Anti-commutator?   Here we outline one possible answer to the question.

The algebra``$\CHOOO$''$\equiv\RCHOOO$ is defined as the left-action maps of $\RCHO$ on itself.  It can easily be seen to generate a faithful representation of the Clifford algebra $\C l(8)$.  Take for example the eight generators 
\begin{equation}\label{8gamma}
\gamma_j \equiv ie_j \hspace{3mm} j=1\dots 6, \hspace{3mm}  \gamma_7 \equiv \epsilon_1 e_7, \hspace{3mm}  \gamma_8 \equiv \epsilon_2 e_7,  \vspace{2mm}
\end{equation}
\noindent where $e_j$ are the standard octonionic imaginary units, and $\epsilon_1,$ $\epsilon_2,$ $\epsilon_3,$ are the standard quaternionic imaginary units.  These generators satisfy the Clifford algebraic anti-commutation relations
\begin{equation} \{  \gamma_n ,\gamma_m \} = 2 \delta_{nm}  \hspace{8mm} n, m = 1\dots 8.
\end{equation}
From these generators we may then define the element
\begin{equation} \gamma_9 \equiv \gamma_1\gamma_2\gamma_3\gamma_4\gamma_5\gamma_6\gamma_7\gamma_8 = \epsilon_3 e_7, 
\end{equation}
\noindent from which we can construct idempotents 
\begin{equation} P_1 \equiv \frac{1}{2}(1-\epsilon_3e_7), \hspace{6mm}  P_2 \equiv \frac{1}{2}(1+\epsilon_3e_7).
\end{equation}
The algebra $\CHOOO$ is subject to an anti-automorphism, $\sim,$ defined by sending $e_i \mapsto -e_i$ and $\epsilon_j \mapsto -\epsilon_j$ for $i=1\dots 7,$ and $j=1\dots 3$, while reversing the order of multiplication.  Additionally, we may then define what we will call the \it hermitian conjugate \rm as $\dagger \equiv*\sim$, where we compose $\sim$ with $*$.  The involution $*$ is the usual complex conjugate that maps the complex imaginary unit $i\mapsto -i$.

\subsection{Multi-actions}

We are now ready to define a generalized multiplication rule, which we will call the $\dagger$\textbf{\textit{-multi-action}}. \rm  The $\dagger$-multi-action is a map from two elements $a,b\in\CHOOO$ to a third element, $m_{\dagger}(a,b)$, which is also in $\CHOOO$.
\begin{equation} \label{sa1}
m_{\dagger}(a,b) \equiv \sum_{i=1}^2 P_ia\hspace{1.1mm}P_ib+P_ib\hspace{1.1mm}a^{\dagger}P_i.
\end{equation}
\noindent Similarly, we will define the $\sim$\textbf{\textit{-multi-action}} \rm  as
\begin{equation} \label{sa2}
m_{\sim}(a,b) \equiv \sum_{i=1}^2 P_ia\hspace{1.1mm}P_ib+P_ib\hspace{1.1mm}\widetilde{a}P_i.
\end{equation}

These multi-actions $m_{\dagger}$ and $m_{\sim}$ generalize the $SU(3)_c$ action demonstrated in this paper, and also may be loosely viewed as a linearization of those actions introduced in Chapter 3.4 of~\cite{thesis}.

These new actions $m_{\dagger}$ and $m_{\sim}$ are called \it multi-\rm actions since they (each) single-handedly split $\CHOOO$ into 
$$\begin{array}{l}
(1) \hspace{3mm}\textup{Lie algebras,} \vspace{3mm}\\

(2)\hspace{3mm} \textup{Jordan algebras, and} \vspace{3mm}\\

(3) \hspace{3mm}\textup{spinor spaces.}  
\end{array}$$

That is, a multi-action supplies the respective multiplication rules to each of these subspaces, simultaneously.  It also supplies the multiplication rules \it between \rm these subspaces.  As readers will see below, $m_{\dagger}$ and $m_{\sim}$ introduce a $\mathbb{Z}_2$ grading, reminiscent of a supersymmetry algebra.  

With the introduction of these multi-actions, we may now identify certain Lie algebras $L_{\dagger}$ and $L_{\sim}$ in $\CHOOO$ as 
\begin{equation} \begin{array}{r}L_{\dagger} \equiv P_1 \hspace{1mm}\CHOOO\hspace{1mm} P_1 + P_2 \hspace{1mm}\CHOOO \hspace{1mm}P_2 \vspace{2mm}
\\-\hspace{1.7mm} \dagger\textup{conjugate}
\end{array}\end{equation}
\noindent using the action $m_{\dagger}$, and
\begin{equation} \begin{array}{r}L_{\sim} \equiv P_1 \hspace{1mm}\CHOOO \hspace{1mm}P_1 + P_2\hspace{1mm} \CHOOO \hspace{1mm}P_2 \vspace{2mm}
\\- \hspace{1.7mm}\sim\textup{conjugate}
\end{array}\end{equation}
\noindent using the action $m_{\sim}$. Similarly, certain Jordan algebras, $J_{\dagger}$ and $J_{\sim}$, may be defined as 
\begin{equation} \begin{array}{r}J_{\dagger} \equiv P_1 \hspace{1mm}\CHOOO\hspace{1mm} P_1 + P_2\hspace{1mm} \CHOOO \hspace{1mm}P_2 \vspace{2mm}
\\+\hspace{1.7mm} \dagger\textup{conjugate}
\end{array}\end{equation}
\noindent using the action $m_{\dagger}$, and
\begin{equation} \begin{array}{r}J_{\sim} \equiv P_1\hspace{1mm} \CHOOO \hspace{1mm}P_1 + P_2\hspace{1mm} \CHOOO\hspace{1mm} P_2 \vspace{2mm}
\\+ \hspace{1.7mm}\sim\textup{conjugate}
\end{array}\end{equation}
\noindent using the action $m_{\sim}$.

Finally, spinor spaces may be identified as
\begin{equation} \begin{array}{r}\Psi \equiv P_1 \hspace{1mm}\CHOOO\hspace{1mm} P_2 + P_2 \hspace{1mm}\CHOOO \hspace{1mm}P_1,
\end{array}\end{equation}
\noindent with the possibility of constructing Majorana representations by adding the complex conjugate.  As vector spaces, it is easy to see that 
\begin{equation}  L_{\dagger} \oplus J_{\dagger} \oplus \Psi = \CHOOO = L_{\sim} \oplus J_{\sim} \oplus \Psi.
\end{equation}
\noindent Clearly, this incorporates the idea of Peirce decomposition of algebras.  These representations will be interesting to study in the context of the derivations, structure algebras, and conformal algebras defined in~\cite{sudbery}.  Also, those familiar with Eilenberg algebras, \cite{boyle}, will readily find close connections to these structures.

Readers are encouraged to verify that  $m_{\dagger}$ and $m_{\sim}$ each map Lie algebra elements $L_i$, Jordan algebra elements $J_j$, and spinors $\Psi_k$, in a fashion which in limited ways mirrors a susy algebra:
\begin{equation}\begin{array}{ll}\label{fashion}
 m(L_i,L_j) = [L_i,L_j]& \in L, \vspace{3mm}

\\ m(J_i,J_j) =\{J_i,J_j\}\hspace{5mm}&\in J, \vspace{3mm}

\\ m(\Psi_i, \Psi_j)+ m(\Psi_j, \Psi_i) = \Psi_j\Psi_i^{\star}+ \Psi_i\Psi_j^{\star}\hspace{5mm}&\in J, \vspace{3mm}

\\ m(\Psi_i, \Psi_j)- m(\Psi_j, \Psi_i) = \Psi_j\Psi_i^{\star}- \Psi_i\Psi_j^{\star}&\in L, \vspace{3mm}

\\m(L_i,\Psi_j) = L_i\Psi_j & \in \Psi, \vspace{3mm}

\\m(J_i,\Psi_j)=J_i\Psi_j& \in \Psi, \vspace{3mm}

\\m(L_i, J_j) = [L_i, J_j] &\in J, \vspace{3mm}

\\m(J_i,L_j) = \{J_i, L_j\} &\in L,\vspace{3mm}

\\m(\Psi_i, L_j) =0, \hspace{5mm}m(\Psi_i,J_j) =0,

\end{array}\end{equation} 
\noindent where $\star$ is meant to represent one of the anti-automorphisms, as the case may be.

In summary, we have just introduced the notion of a \it multi-action. \rm  A multi-action can be thought of, intuitively, as a machine which splits and processes our original algebra into Lie algebras, Jordan algebras, and spinor spaces.  When this machine is given a pair of elements from the Lie algebraic subspace, it processes them via a commutator, so as to output another Lie algebraic element.  When it is fed a pair of elements from the Jordan algebraic subspace, it processes them via an anti-commutator, so as to output another Jordan algebraic element.  When this machine is fed a Lie algebraic element and a spinor, it can output a linearized transformation on that spinor, and so on.   In short, a multi-action can be thought of as a generalized multiplication rule, which simultaneously orchestrates the behaviour of a collection of algebraic objects, each familiar to physics.   It is with the $\CHOOO$ multi-actions $m_{\dagger}$ and $m_{\sim}$, or some variation thereof, that we hope to uncover the group representation structure of the standard model.

\subsection{Preliminary findings}

Now when constructing a field theory, we might hope to associate these fields with certain special invariant subspaces of $\RCHOOO$.  (See \cite{thesis} section 2.3.)  Specifically,  
\begin{equation}\begin{array}{cl}\label{real}
\Psi + \Psi^*\hspace{1mm} &\sim\hspace{2mm} \textup{fermions,} \vspace{2mm}\\

J\hspace{2mm} &\sim\hspace{2mm} \textup{gauge bosons,} 
\end{array}\end{equation}
\noindent with field strengths to be built up from gauge bosons in the usual way.   We propose associating gauge bosons with elements of a Jordan algebra based on the observation that certain Jordan algebras can generalize the $2\times 2$ hermitian matrices,  much in the way that $A^a_{\mu}\gamma^{\mu}\Lambda_a$ generalizes $p_{\mu}\sigma^{\mu}$ from the Weyl equation.  Of course, care must be taken to ensure that our new gauge fields indeed transform correctly.

It is clear from this construction that we have more than one Jordan algebra to choose from.  One natural option may then be to consider the intersection of the Jordan algebras we have available.  Let us then consider the Jordan algebra $J_{\dagger} \cap J_{\sim}$.  Upon calculation, readers will find that this algebra $J_{\dagger} \cap J_{\sim}$ is $28 + 28$ $\R$-dimensional. On the other hand, the Lie algebra $L_{\dagger} \cap L_{\sim}$ is $36 + 36$ $\R$-dimensional, and appears to be given by two copies of the symplectic algebra $sp(8, \R)$.  These details, among many in this addendum, are yet to be more carefully confirmed.

Now incidentally, $\sim$ and $\dagger$ are not the only anti-automorphisms of $\RCHOOO.$  We furthermore have $\rho,$ the well-known reversal anti-automorphism of $\C l(8)$.  This reversal anti-automorphism sends any vector in the Clifford algebra's generating space to itself, and likewise maps the identity element to itself, \cite{budi}.  Beyond reversal, we may build a further anti-automorphism by composing reversal, $\rho,$ together with the automorphism of grade involution, $\alpha$.  Grade involution is an automorphism which maps the $\C l(8)$ generators $\gamma_{\nu} \mapsto -\gamma_{\nu},$ and leaves the identity element invariant.

With these new anti-automorphisms, $\rho$ and $\alpha \rho$, we may easily construct analogues of the multi-actions~(\ref{sa1}) and (\ref{sa2}). From here, it is then straightforward to find a broader range of Jordan and Lie algebras: $J_{\rho},$  $L_{\rho},$ $J_{\alpha\rho},$ $L_{\alpha\rho},$ $J_{\rho *},$  $L_{\rho *},$ etc.  These Lie and Jordan algebras may then in turn be used to construct further intersection algebras.  For example, we find that $L_{\rho}\cap L_{\alpha \rho} \cap L_{\rho *}$ is $28 + 28$ $\R$-dimensional, and appears to be given by two copies of the $so(8)$ Lie algebra.

The most tightly constrained Jordan and Lie intersection algebras we have at this point may be expressed as $J_{\rho}\cap J_{\alpha \rho} \cap J_{\rho *}\cap J_{\sim}$ and $L_{\rho}\cap L_{\alpha \rho} \cap L_{\rho *}\cap L_{\sim}$ (or equivalently, $J_{\rho}\cap J_{\alpha \rho} \cap J_{\dagger}\cap J_{\sim}$ and $L_{\rho}\cap L_{\alpha \rho} \cap L_{\dagger}\cap L_{\sim}$, or ...).  We find that both the Jordan and Lie intersection algebras are $16 + 16$ $\R$-dimensional, and the Lie algebra appears to be given by two copies of $u(4)$.  In fact, this final Lie algebra appears to embody the more general identity $sp(2n, \R) \cap so(2n) = u(n).$  Clearly, this $u(4)$ Lie algebra could in principle include at least the standard model's unbroken gauge symmetry $su(3)_c$.  However, we are yet to find an exact match to the standard model's full set of gauge symmetries.  

One option under investigation is to consider not just the left action of $\RCHO$ on itself, but rather both the left and right actions.  The chain algebra in this case generates a faithful representation of the larger Clifford algebra $\C l(10).$  From here we may attempt to generalize our Lie algebra structure to $sp(10, \R) \cap so(10) = u(5).$  Once $u(5)$ is obtained, only one simple step is subsequently required so as to obtain the standard model's $su(3)_C\oplus su(2)_L \oplus u(1)_Y$, together with an additional $u(1)_{X},$ as explained in~\cite{aguts}, \cite{malala}, and \cite{trials}.

With the multi-actions and the intersection algebras now defined, we will demonstrate precisely why these proposals deserve further investigation.

Let us consider the generators $i\Lambda_j$ of $SU(3)\subset Aut(\mathbb{O})$, represented as in equations (\ref{g2}).  Readers may confirm that the $i\Lambda_j$ reside in the Lie algebra $L_{\rho}\cap L_{\alpha \rho} \cap L_{\rho *}\cap L_{\sim}.$  Inputting $a=i\Lambda_j$ and $b\in\CHOOO$ into multi-actions~(\ref{sa1}), (\ref{sa2}), or their analogues, will clearly lead to the same result.  
\begin{equation}\begin{array}{l}\label{multi3}
m(\hspace{1mm}i\Lambda_j,\hspace{1mm}\CHOOO\hspace{1mm}) = \vspace{2mm} 
\\  \hspace{3mm}\sum_{i=1}^2 \hspace{2mm}P_i\hspace{1mm}i\Lambda_j\hspace{.6mm}P_i\hspace{1mm}\CHOOO-P_i\hspace{1mm}\CHOOO\hspace{1.3mm}i\Lambda_j P_i.
\end{array}\end{equation}
Under this action, we find the breakdown of $\CHOOO$ into $SU(3)$ representations as follows:
\begin{equation}\begin{array}{lccccc} \label{bd} \CHOOO \mapsto \vspace{2mm}
\\ \hspace{8mm} (4 \times \mathbf{\underline{8}})&\oplus &\hspace{1mm} (24\times  \mathbf{\underline{3}}) &\oplus &(18\times \mathbf{\underline{1}})& \oplus\vspace{2mm}
\\  \hspace{8mm}(4 \times \mathbf{\underline{1}}) &\oplus &\hspace{1mm} (24\times  \mathbf{\underline{3}^*}) &\oplus &(18\times \mathbf{\underline{1}})& \oplus \vspace{2mm} 
\\ \hspace{8mm}(4 \times \mathbf{\underline{1}})& \oplus&\mathcal{C}_{36},\hspace{4mm}
\end{array}\end{equation}
\noindent where $ \mathcal{C}_{36}$ is a $36$ $\C$-dimensional space, to be discussed shortly.  It should be noted that the representations listed above are over the field $\C$, which would seem to introduce a doubling in the number of degrees of freedom, relative to what one might expect of standard model states.  However, this may be addressed by considering those subspaces invariant under complex conjugation $i\mapsto -i$, as proposed in relations (\ref{real}), and the paragraphs that follow them,~\cite{thesis}.

Clearly, this breakdown~(\ref{bd}) comes quite close to those $SU(3)_c$ representations we find in the standard model, now with the capacity to accommodate spin and polarization.  To be specific, the first 220 degrees of freedom described in this decomposition do indeed exhibit $SU(3)_c$ behaviour we see in elementary particle physics.  If we were to construct the standard model by starting out with four real degrees of freedom allocated for the spacetime indices of each gauge boson, and four real degrees of freedom per spinor, then the above states could (for example) account for eight gluons, one photon, one Z boson, three generations of left- and right-handed quarks, three generations of left-handed leptons, and three generations of right-handed charged leptons.  Still left to describe would be W bosons, the Higgs, and potentially right-handed neutrinos. 

Now the remaining available space, $\mathcal{C}_{36}$, is something of a curiosity.  Under the action (\ref{multi3}), we find that $\mathcal{C}_{36}$ behaves as two copies of $\mathbf{\underline{3}} \otimes\mathbf{\underline{3}} = \mathbf{\underline{6}}\oplus \mathbf{\underline{3}^*}$, plus its complex conjugate.  In corroboration with the recent work of Gording and Schmidt-May,~\cite{gsm}, we find that it might be possible to alleviate extraneous-$su(3)_c$ tension by having such states act on $\Psi$ from the right hand side, instead of the left.  In this case, triplet transitions may be identified as transitions between generations instead of colour.  Said another way, one may seek out special conditions under which $\Psi$ may be reinterpreted as $\Psi^{\star}$.  Clearly this topic will be  subject to further investigation.

\smallskip

\begin{acknowledgments}
This work has been partially supported by STFC consolidated grant ST/P000681/1.  The author is furthermore grateful for support from the NSERC Postdoctoral Fellowship, and from the Walter Grant Scott Research Fellowship in Physics at Trinity Hall, University of Cambridge.  A special thanks goes in particular to my friends and colleagues for discussions and their feedback on this paper:   Latham Boyle, Brage Gording, Judd Harrison, Mia Hughes, Alessio Marrani, Mike Rios, Christopher Thomas, Ivan Todorov, and David Tong.



\end{acknowledgments}

\medskip

\end{document}